# Visualization of one-dimensional diffusion and spontaneous segregation of hydrogen in single crystals of VO$_2$

T. Serkan Kasırga [1,2], Jim M. Coy[1], Jae H. Park[1] and David. H. Cobden[1]

[1]Department of Physics, University of Washington, Seattle, WA 98195

[2]UNAM−Institute of Materials Science and Nanotechnology, Bilkent University, Ankara 06800, Turkey

**ABSTRACT**

**Hydrogen intercalation in solids is common, complicated, and very difficult to monitor. In a new approach to the problem, we have studied the profile of hydrogen diffusion in single-crystal nanobeams and plates of VO$_2$, exploiting the fact that hydrogen doping in this material leads to visible darkening near room temperature connected with the metal-insulator transition at 65 °C. We observe hydrogen diffusion along the rutile c-axis but not perpendicular to it, making this a highly one-dimensional diffusion system. We obtain an activated diffusion coefficient, $\sim 0.01 \, e^{-0.6 \, eV/k_B T}$ cm$^2$sec$^{-1}$, applicable in metallic phase. In addition, we observe dramatic supercooling of the hydrogen-induced metallic phase and spontaneous segregation of the hydrogen into stripes implying that the diffusion process is highly nonlinear, even in the absence of defects. Similar complications may occur in hydrogen motion in other materials but are not revealed by conventional measurement techniques.**

**TEXT**

Hydrogen dissolves in many solids, with effects on their physical properties which are relevant to storage[1], catalysis, sensing[2], and material degradation[3]. Understanding how hydrogen moves within a solid is thus important, but is very challenging because of the combination of the difficulty of detecting hydrogen, the complexity of the transport processes, and the sensitivity of kinetics and energetics to microscopic nonuniformity and defects[4]. It has traditionally been studied, predominantly in metals and semiconductors, by neutron scattering[5], nuclear[6] and muon magnetic resonance[7], and other techniques which do not yield spatial profiles. It is thus hard to know when deviations from a simple Fick's law behavior are relevant, such as when diffusion is guided by dislocations, grain boundaries or strain fields associated with twinning or structural modifications, or if there is spontaneous segregation. Recently, information on the spatial profile has been obtained using the change in color on hydrogenation of thin yttrium[8] and vanadium layers[9]. In these cases the behavior was explained using Fick's law, though some anomalies were present.

Hydrogen can also diffuse readily in some oxides[10], including rutile $TiO_2$ and $VO_2$ [11]. The rutile structure contains oxygen-lined channels parallel to the tetragonal c-axis along which hydrogen can hop between temporary O-H bonds (Fig. 1a). Diffusion perpendicular to these channels requires the protons to move between densely spaced metal ions, and has a much higher energy barrier[12,13]. Pure, unstrained rutile-structured (R) $VO_2$ has a well-known first-order metal-insulator transition (MIT) to a monoclinic insulating phase (M1) on cooling through $T_c = 65$ °C[14,15]. Interestingly, it has recently been found[16,17] that hydrogen in $VO_2$ reduces the gap in the insulating phase while slightly reducing $T_c$, and the material becomes fully metallic above some concentration. Consequently, a thin $VO_2$ sheet on a suitable substrate appears darker when hydrogenated, making it possible to directly image hydrogen motion in this oxide. In addition, small $VO_2$ nanobeams and plates offer the unusual opportunity to study diffusion in an unstrained, domain-free single crystal.

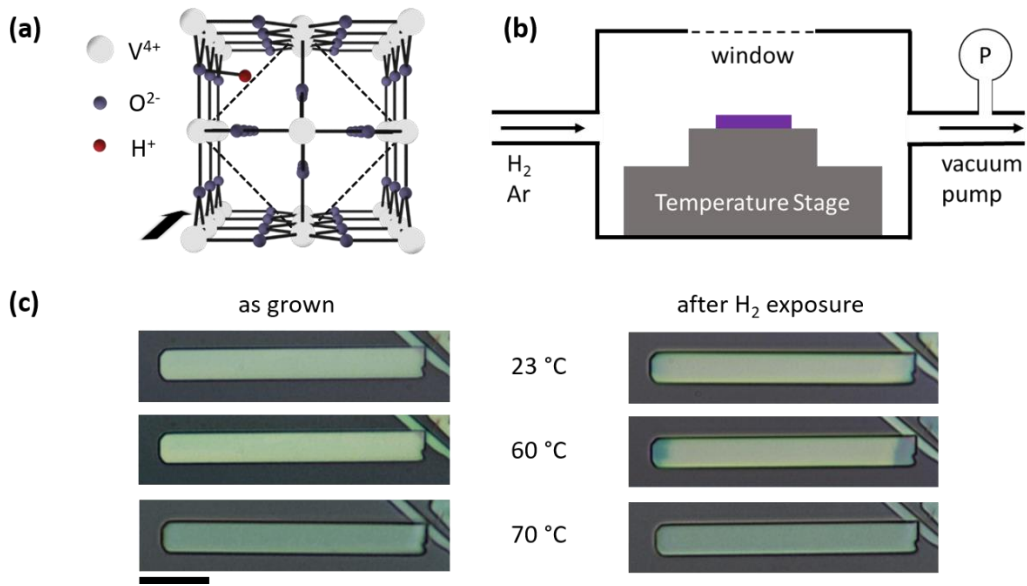

**Figure 1.** (a) View along the rutile c-axis shows the channels lined by oxygen atoms and a proton bound to one. Dashed line shows the base of the rutile unit cell and the black arrow indicates the direction of the rutile c-axis. (b) Schematic of the hydrogenation chamber, hermetically sealed with gas inlets and pump outlet. Gas flow is controlled by mass flow controllers and the chamber pressure is monitored by pressure sensors. The stage can be varied from -40 °C to 200°C. (c) Optical images of a $VO_2$ nanobeam of thickness about 100 nm, before and after hydrogen exposure (15 minutes of 5 ml/min $H_2$ and 50 mL/min Ar at 50 mBar at 100 °C). Scale bar is 10 µm.

To exploit this idea we prepared strain-free $VO_2$ crystals and observed their darkening on exposure to hydrogen gas, using a palladium catalyst to dissociate the hydrogen on the surface. We thereby successfully determined the diffusion coefficient along the c-axis and its activation energy. We found that it is orders of magnitude greater than transverse to the channel axis, as in rutile $TiO_2$, making the diffusion one-dimensional. More surprisingly, we observed extreme supercooling of the metallic state, and spontaneous segregation of the hydrogen into stripes during diffusion. The latter implies that the diffusion process deviates strongly from Fick's law - a fact that would be disguised in conventional measurements. As a result, it appears impossible to achieve stable uniform moderate concentrations of hydrogen in $VO_2$, as would be desirable for tuning the MIT for applications. We postulate that the segregation is related to the recently reported[18] structural change from monoclinic (M1) to orthorhombic (O1 and O2) occurring at moderate H doping levels, with elastic energies favoring a domain structure. Such segregation may be common in other solids (hydrogen uptake is typically accompanied by changes in the equilibrium structure) but has not been noticed because of the lack of probes.

We grow the crystals by physical vapor transport on oxidized silicon wafers[19]. They are usually elongated along the rutile c-axis, and when released from the substrate they are optically uniform and exhibit sharp MITs at 65 °C implying a lack of impurities or twinning[15,20,21]. To counter strain caused by substrate adhesion[22], the crystals are either transferred to a soft polymer, polydimethylsiloxane (PDMS), or cantilevered from the edge of an oxidized silicon chip. To transfer the crystals on to PDMS, a PDMS coated chip is pressed against the growth substrate. We dip the substrate briefly dipped into buffered oxide etchant solution to loosen the crystals for a higher yield transfer and to remove any oxide formation on the surface of the crystals. 1 nm of Pd is evaporated to catalyze dissociation of $H_2$ on the $VO_2$ surface[23] (see suppl. mater. for further details). They are then exposed to hydrogen carried by argon flowing through a sealed chamber as indicated in Fig. 1(b).

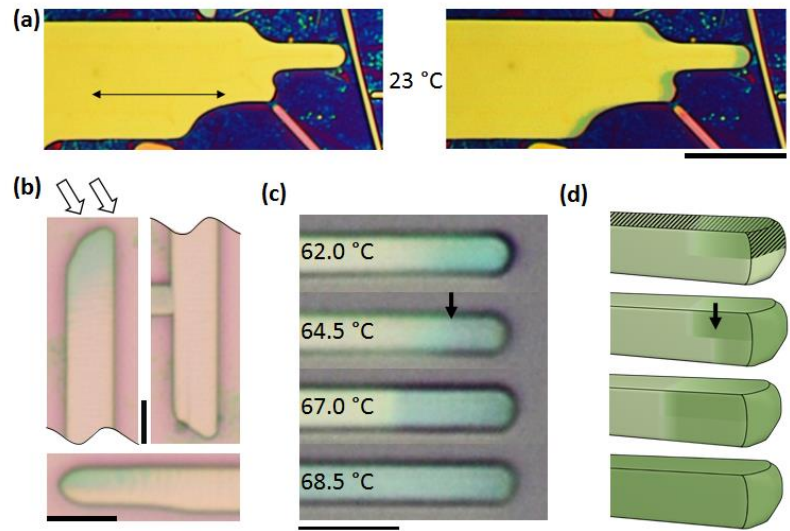

**Figure 2.** (a) Images of an irregularly shaped plate crystal at 23 °C before (left) and after (right) hydrogen exposure, clearly manifesting one-dimensional diffusion along the channel axis (arrow). Scale bar is 10 µm. (b) Images of crystals with oblique Pd evaporation (direction indicated by the arrow) after hydrogen exposure. The top two images are the ends of the same nanobeam. Darkening only occurs from the coated edges, demonstrating the importance of the Pd catalyst for hydrogen injection. (c) Images of a nanobeam on PDMS after hydrogen exposure. The arrow indicates a boundary between two regions with different darkening. (d) Possible explanation for this. Hydrogen enters mainly the upper side of the end where Pd is present (indicated by hatching), leading to a lower concentration in the underneath side adjacent to the substrate. Scale bars are 5 µm.

Figure 1(c) shows optical images of a VO$_2$ nanobeam on PDMS at several temperature, before (left) and after (right) an exposure to H$_2$ at 100 °C (thus in the R phase). At 70 °C, before exposure the nanobeam is fully metallic, and after exposure it looks exactly the same. At 23 °C, before exposure it is fully insulating (paler color), and after exposure a faint darkening is visible near the ends. At 60 °C, before exposure it is still fully insulating, but after exposure there are distinct regions at the ends having the full metallic darkness. We directly infer that hydrogen enters from at the ends, moving along the c-axis on a scale of several microns and producing a concentration gradient so that both the insulator gap and $T_c$ are reduced near the ends. A lack of darkening far from the ends indicates that negligible hydrogen enters from the top surface, which is also covered with catalyst, in spite of the much shorter distance involved (~50 nm) for diffusion perpendicular the c-axis.

The extreme anisotropy of the diffusion is manifested in many ways. For one example, in crystals with complicated shapes, such as the one shown in Fig. 2a, the edge of the metallized region after exposure has the shape of the edge of the crystal translated along the c-axis. For another example, if we evaporate the Pd at a shallow oblique angle so that crystal side walls facing away from the source are not coated with Pd, we do not see darkening near these sides (Fig. 2b). This also demonstrates the importance of the Pd catalyst.

Another common observation is illustrated in Fig. 2c. In this nanobeam while warming we see distinct regions with different levels of darkening. This could be partly explained by a difference in color between doped insulator and the metallic phase, but another factor is in play. The end of a nanobeam is often pyramidal, and when the Pd is evaporated from above only the upper part of the end is covered, as sketched in Fig. 2d. Hence hydrogen only enters the upper part of the crystal, which thus darkens further from the end at a given temperature.

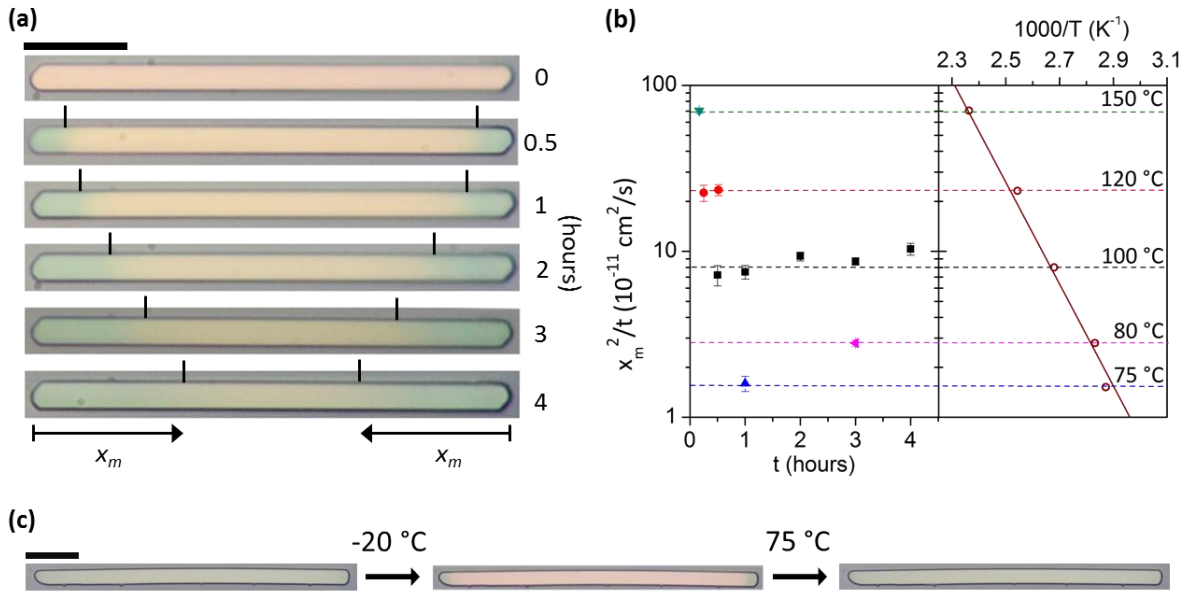

**Figure 3.** (a) Series of images taken at 60 °C before and after hydrogen exposure at 100 °C for the time $t$ indicated on the right. Ticks shows the position $x_m$ where the darkening is half that at the ends. All scale bars are 10 µm. (b) Plot for extracting the diffusion coefficient $D = x_m^2/t$ at several temperatures. The Arrhenius plot on the right of the graph yields

activation energy $E_a = 0.6$ eV. **(c)** Supercooling seen at longer exposures (2 hours at 100 °C). All images are at room temperature.

Fortunately, some crystals on PDMS exhibited behavior consistent with uniform hydrogen injection at the ends at moderate exposure levels. Figure 3a shows a series of images of such a nanobeam after exposure to an H$_2$:Ar (1:10) gas mixture at 100 °C for different lengths of time $t$. They are taken at 60 °C, because the darkening is greater than at room temperature yet the metal-insulator boundary is not as sharp as it is near $T_C$. We see a darkness gradient related to the local hydrogen concentration resembling a simple 1D diffusion profile. From Fick's law in one dimension, $\partial n/\partial t = D\partial^2 n/\partial x^2$, with diffusion coefficient $D$, assuming zero concentration $n = 0$ for time $t < 0$ and a single boundary at $x = 0$ at which $n$ is held at a constant value $n_0$ for $t > 0$, we have $n(x,t) = n_0 \text{erfc}(\frac{x}{2\sqrt{Dt}})$. According to this equation, after time $t$ the concentration decays by a half at distance $x = x_m$ where $x_m/2\sqrt{Dt} = \text{erfc}^{-1} 1/2 \approx 1/2$, or $x_m^2/t \approx D$.

To estimate $x_m$, the diffusion length, in the experiments we measure the positions at which the darkening is half that at the end, indicated by tick marks in Fig. 3a. In Fig. 3b we plot $x_m^2/t$ for a series of exposure times $t$ at 100 °C (black). The results are consistent with the Fick's law prediction and yield $D = (8 \pm 1) \times 10^{-11}$ cm$^2$s$^{-1}$. We estimate the error here taking into account the fact that the darkness is an unknown nonlinear function of the hydrogen concentration. Even if there is for example a threshold value for hydrogen concentration to produce a color change, the effect on the estimated diffusion length will be minor since the temperature dependence of the position of the boundary is what matters. We also plot values obtained using other exposure temperatures $T$. An Arrhenius fit, shown on the right, yields an activation energy of $E_a = 0.6 \pm 0.1$ eV, similar to that reported for diffusion along the c-axis of VO$_2$[24] and TiO$_2$[12,13]. Combining the results we find $D(T) \approx 0.01\, e^{-0.6/k_B T}$ cm$^2$s$^{-1}$ along the rutile c-axis. These values are also consistent with the recent theoretical studies[25,26].

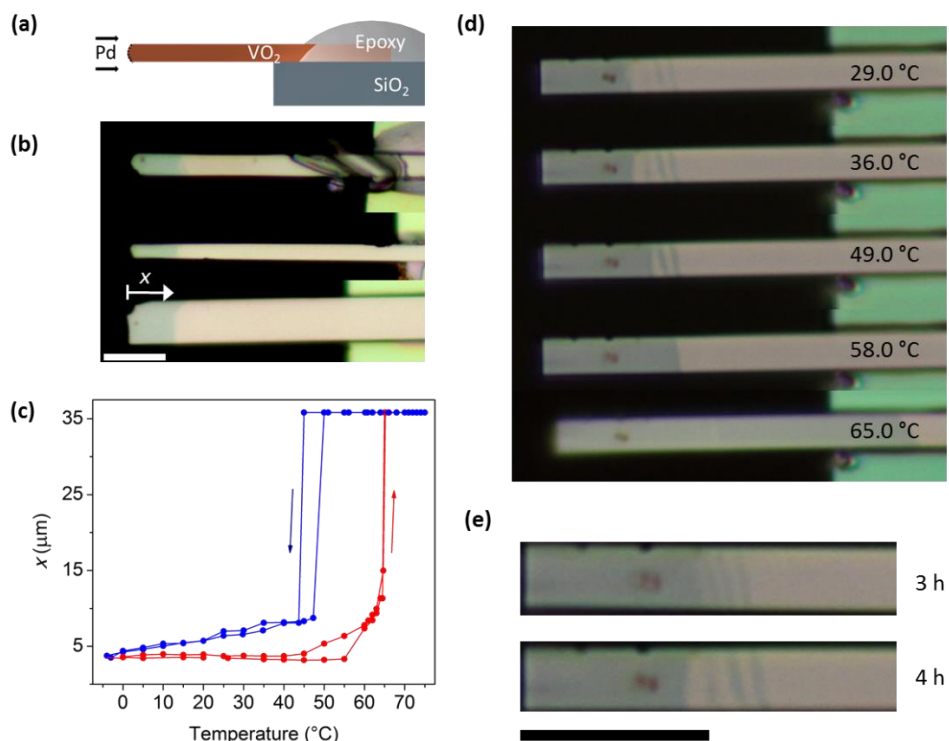

**Figure 4. (a)** Nanobeams are glued down cantilevered from the edge of a silicon chip and Pd catalyst is evaporated on the free ends. **(b)** Images at 55 °C of three cantilevered nanobeams with different cross-sections after exposure to hydrogen for 20 minutes at 120 °C. All show the same metallized length $x$. **(c)** Measurement of $x$ vs. temperature for one cantilever, cycled twice to the fully metallic state at 60 °C. **(d)** Images at a series of temperatures of a cantilevered nanobeam after hydrogen exposure at 90 °C for 4 hours. Alternating darker and lighter stripes can be below 50 °C. **(e)** The period of the stripes increases with exposure time. Images are taken at 36 °C. Scale bars are all 10 μm.

At longer exposures (lower images in Fig. 3a), a complicating supercooling effect is present. Immediately after cooling from the fully metallic state (at 70 °C) to room temperature the crystals remain fully metallic. Only if left overnight, or cooled further, do they return to the insulating state with darkened ends as in the lower images in Fig. 3a. After 4h at 100 °C a dip in liquid nitrogen is needed to recover the insulating state. An example is shown in Figure 3c.

To eliminate any residual strain effects due to the PDMS substrate, we studied cantilevered nanobeams transferred using a micromanipulator onto the edge of an oxidized silicon chip and fixed with epoxy as sketched in Fig. 4a. These were rinsed in dilute HF to remove surface $V_2O_5$ prior to evaporating 1 nm of Pd directed at the ends. Fig. 4b shows three nanobeams of different size after hydrogen exposure. A metallic region can be seen with an interface at distance $x$ from the end. Fig. 4c shows the variation of $x$ on cycling the temperature up (red) and down (blue). It is reproducible over two cycles, and almost identical for all three nanobeams, implying that this is intrinsic defect-free single-crystal behavior. In addition to the usual hysteresis in position of the large jump at the MIT, there is large hysteresis in the interface position implying that it is "sticky" and its motion incurs a penalty of dissipated energy.

Most strikingly, our attempts to produce a uniform hydrogen concentration failed. After a long exposure (4h at 90 °C), rather than seeing the darkening spread smoothly along the cantilevers we observed dark stripes which are faint but visible at room temperature, as shown in Fig. 4d. On warming these become sharper, and then the paler regions between them turn dark too, but the insulating region beyond the last stripe does not darken until close to 65 °C implying that it remains undoped. In addition, the period of the stripes increases, and each stripe propagates along the nanobeam and becomes darker, with increasing exposure time, as shown in Fig. 4e.

The formation of stripes implies that the hydrogen tends to self-segregate in single-crystal $VO_2$ (see supplementary materials for further evidence), and thus that there are large nonlinear diffusion forces that violate Fick's law. From the reproducibility between cantilevered nanobeams, the role of defects[4] and external strain[27] can be eliminated. Instead, the behavior can be explained qualitatively as follows: above a certain H concentration a structural distortion of the R phase becomes favorable to an orthorhombic O phase[18] and O domains appear with sizes that minimize the elastic energy cost of creating O-R interfaces. Thereafter, hydrogen tends to move from R to O regions to further reduce the total free energy. Combined with the flow of hydrogen in from the free end this can produce evolving stripes of O phase. It seems likely that the stickiness of the metal-insulator interface and the supercooling described above are also related to the competition between different displacive structures combined with elastic energy and diffusion.

In summary, our studies of hydrogen diffusion in single crystals of $VO_2$ by optical microscopy have revealed extreme anisotropy and highly nonlinear effects in diffusion that may be relevant in many other materials.

**REFERENCES**

1 Schlapbach, L.; Züttel, A. Hydrogen-storage materials for mobile applications. *Nature* **2001,** *414*, 353-358.

2 Hübert, T.; Boon-Brett, L.; Black, G.; Banach, U. Hydrogen sensors- A review. *Sensors and Actuators B* **2011**, *157*, 329-352.

3 Hirth, J.P. Effects of hydrogen on the properties of iron and steel. *Metall. Trans. A* **1980**, *11*, 861-890.

4 Myers, S.M.; Baskes, M.I.; Birnbaum, H.K.; Corbett, J.W.; DeLeo, G.G.; Estreicher, S.K.; Haller, E.E.; Jena, P.; Johnson, N.M.; Kirchheim, R.; Pearton, S.J.; Stavola, M.J. Hydrogen interactions with defects in crystalline solids. *Rev. Mod. Phys.* **1992**, *64*, 559.

5 Rowe, J.M.; Sköld, K.; Flotow, H.E.; Rush, J.J. Quasielastic neutron scattering by hydrogen in the α and β phases of vanadium hydride. *J. Phys. Chem. Sol.* **1971**, *32*, 41-54.

6 Hofmann, D.M.; Hofstaetter, A.; Leiter, F.; Zhou, H.; Henecker, F.; Meyer, B.K.; Orlinskii, S.B.; Schmidt, J.; Baranov, P.G. Hydrogen: A relevant shallow donor in Zinc Oxide. *Phys. Rev. Lett.* **2002**, *88*, 045504.

7 Kreitzman, S.R.; Hitti, B.; Lichti, R.L.; Estle, T.L.; Chow, K.H. Muon-spin-resonance study of muonium dynamics in Si and its relevance to hydrogen. *Phys. Rev. B* **1995**, *51*, 13117.

8 den Broeder, F.J.A.; van der Molen, S.J.; Kremers, M.; Huiberts, J.N.; Nagengast, D.G.; van Gogh, A.T.M.; Huisman, W.H.; Koeman, N.J.; Dam, B.; Rector, J.H.; Plota, S.; Hanzen, R.M.N.; Jungblut, R.M.; Duine, P.A.; Griessen, R. Visualization of hydrogen migration in solids using switchable mirrors. *Nature* **1998**, *394*, 656-658.

9 Palsson, G.K.; Bliersbach, A.; Wolff, M.; Zamani, A.; Hjörvarsson, B. Using light transmission to watch hydrogen diffuse. *Nature Comm.* **2012**, *3*, 892.

10 Fripiat, J. J.; Lin, X. Hydrogen intercalation within transition metal oxides: entropy, enthalpy, and charge transfer. *J. Phys. Chem.* **1991**, *96,* 1437-1444.

11 Chippindale, A.M.; Dickens, P.G.; Powell, A.V. Synthesis, characterization, and inelastic neutron scattering study of hydrogen insertion compounds of $VO_2$ (rutile). *J. Solid State Chem.* **1991**, *93*, 526-533.

12 Johnson, O.W.; Paek, S.-H.; DeFord, J.W. Diffusion of H and D in $TiO_2$: Suppression of internal fields by isotope exchange. *J. Appl. Phys.* **1975**, *46*, 1026.

13 Bates, J.B.; Wang, J.C.; Perkins, R.A. Mechanisms for hydrogen diffusion in $TiO_2$. *Phys. Rev. B* **1979**, *19*, 4130-4139.

14 Morin, F.J. Oxides which show a metal-to-insulator transition at the Neel temperature. *Phys. Rev. Lett.* **1959**, *3,* 34.

15 Park, J. H.; Coy, J. M.; Kasirga, T. S.; Huang, C.; Fei, Z.; Hunter, S.; Cobden, D. H. Measurement of a sold-state triple point at the metal-insulator transition in $VO_2$. *Nature* **2013**, *500*, 431-434.

16 Andreev, V.; Kapralova, V.; Klimov, V. Effect of hydrogenation on the metal-semiconductor phase transition in vanadium dioxide thin films. *Phys. Solid State* **2007**, *49*, 2318-2322.

17 Wei, J.; Ji, H.; Guo, W. H.; Nevidomskyy, A. H.; Natelson, D. Hydrogen stabilization of metallic vanadium dioxide in single-crystal nanobeams. *Nature Nanotechnol.* **2012, 7**, 357-362.

18 Filinchuk, Y.; Tumanov, N.A.; Ban, V.; Ji, H.; Wei, J.; Swift, M.W.; Nevidomskyy, A.H.; Natelson, D. In Situ Diffraction Study of Catalytic Hydrogenation of $VO_2$: Stable Phases and Origins of Metallicity. *J. Am. Chem. Soc.* **2014**, *136*, 8100-8109.

19 Guiton, B.S.; Gu, Q.; Prieto, A.L.; Gudiksen, M.S.; Park, H. Single-crystalline vanadium dioxide nanowires with rectangular cross sections. *J. Am. Chem. Soc.* **2005**, *127*, 498.

20 Cao, J.; Gu, Y.; Fan, W.; Chen, L.Q.; Ogletree, D.F.; Chen, K.; Tamura, N.; Kunz, M.; Barret, C.; Seidel, J.; Wu, J. Extended mapping and exploration of the vanadium dioxide stress-temperature phase diagram. *Nano Lett.* **2010**, *10*, 2667-2673.

21 Wei, J.; Wang, Z.; Chen, W.; Cobden, D.H. New aspects of the metal-insulator transition in single-domain vanadium dioxide nanobeams. *Nature Nanotechnol.* **2009**, *4*, 420.

22 Wu, J.; Gu, Q.; Guiton, B.S.; de Leon, N.P.; Ouyang, L.; Park, H. Strain-induced self organization of metal-insulator domains in single crystalline VO$_2$ nanobeams. *Nano Lett.* **2006**, *6*, 2313-2317.

23 Sermon, P.A.; Bond, G.C. Hydrogen spillover. *Catal. Rev.* **1973**, *8*, 211-239.

24 Lin, J.; Ji, H.; Swift, M. W.; Hardy, W. J.; Peng, Z.; Fan, X.; Nevidomskyy, A. H.; Tour, J. M.; Natelson, D. Hydrogen diffusion and stabilisation in single-crystal VO$_2$ micro/nanobeams by direct atomic hydrogenation. *Nano Lett.* **2014**, 14, 5445-5451.

25 Warnick, K. H.; Wang, B.; Pantelides, S. T. Hydrogen dynamics and metallic phase stabilization in VO$_2$. *Appl. Phys. Lett.* 2014, **104**, 101913.

26 Cui, Y.; Shi, S.; Chen, L.; Luo, H.; Gao, Y. Hydrogen-doping induced reduction in the phase transition temperature of VO$_2$: a first-principles study.

27 Gorsky, W.S. *Phys. Z. Sowjetunion* **1935**, *8*, 457.

# Visualization of one-dimensional diffusion and spontaneous segregation of hydrogen in single crystals of VO$_2$: Supporting Information

T. Serkan Kasırga [1,2*], Jim M. Coy[1], Jae H. Park[1] and David. H. Cobden[1]

[1]Department of Physics, University of Washington, Seattle, WA 98195

[2]UNAM−Institute of Materials Science and Nanotechnology, Bilkent University, Ankara 06800, Turkey

## 1. Hydrogen Doping Setup

We built a hydrogen doping setup to visualize the effects of hydrogen on single crystals of VO$_2$. The main body of the setup is the hydrogenation chamber shown in Fig. S1a. The chamber contains two gas inlets, one for hydrogen and the other for argon, one gas outlet, water cooling tubing for the thermoelectric plate and a 9-port electrical connector. The chamber is hermetically sealed and can reach to 10$^{-7}$ mBar with a turbo-molecular pump. A special thermoelectric plate is chosen to be able to heat the samples up to 200 °C. The lowest stable temperature we can reach with the temperature stage is -40 °C. The stage temperature is measured with a Pt-100 sensor calibrated against the melting points gallium, palmitic acid and the metal-insulator transition temperature[1] of VO$_2$ and it is controlled with a closed-loop temperature controller. When the chamber is sealed, mK stability is achievable. Two mass flow controllers are used to measure the gas flow rates and adjust the mixing ratio of argon and hydrogen. Monitoring the pressure in the chamber is done by two pressure gauges, before and after the chamber. Gas flow rates, chamber pressure and the stage temperature is controlled via a LabView program.

[*] To whom correspondence should be addressed: kasirga@unam.bilkent.edu.tr

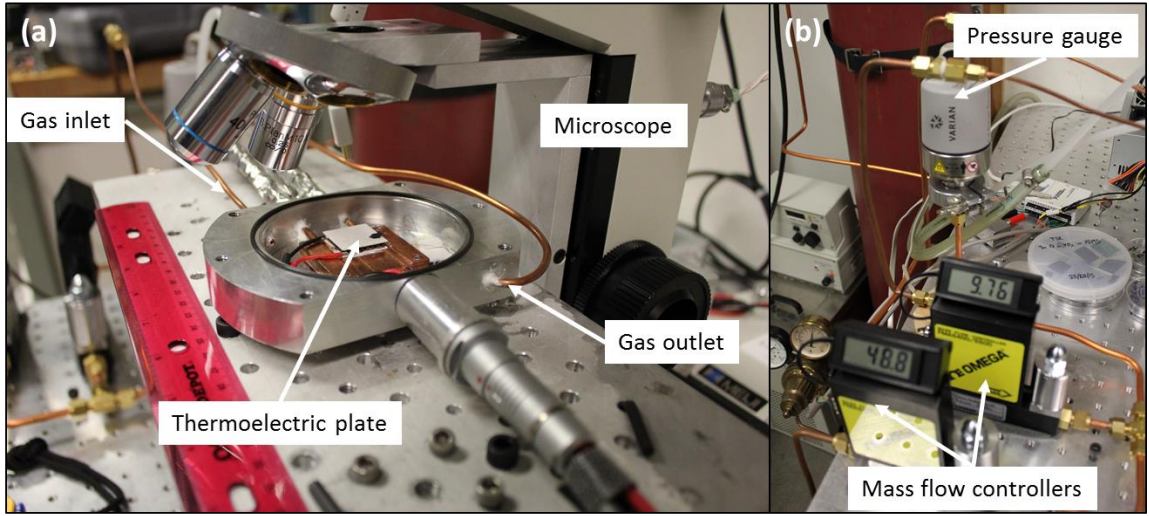

**Figure S1.** Pictures of the hydrogen doping setup are shown. **(a)** Close-up view of the custom-made doping chamber. The chamber can be hermetically sealed with a windowed lid (not shown in the picture). **(b)** Mass flow controllers and the outlet pressure gauge is shown. There is another pressure gauge on the gas inlet port.

## 2. Sample Preparation

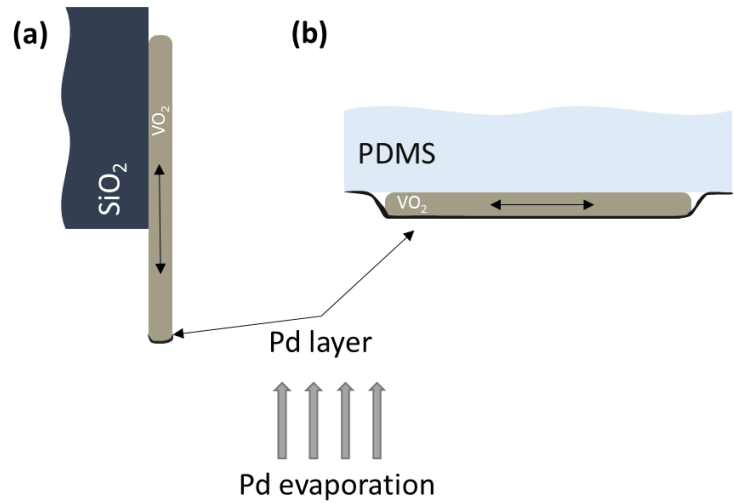

**Figure S2. (a)** Schematic illustration of cantilevered $VO_2$ nanobeam on $SiO_2$ evaporated with Pd straight from below and **(b)** on PDMS substrate. Exaggerated Pd layer illustrates the fact that when Pd evaporated from below part of the faces opening to the crystals c-axis (indicated by double headed arrows) is not covered with Pd.

Strain effects play a significant role on the physical properties of $VO_2$. To eliminate any complications that might arise due to strain effects in hydrogen doped $VO_2$, we prepared strain-free $VO_2$ crystals by either cantilevering the nanobeams from the edge of a silicon chip or transferring onto a soft substrate, polydimethylsiloxane (PDMS).

To prepare the cantilevered samples, nanobeams are loosened from the $SiO_2$ substrate, where they have grown on initially, with a buffered oxide etchant (BOE) solution for about a minute. Then, suitable nanobeams are selected and picked-up using a sharp probe with the help of electrostatics, attached to a nanomanipulator. Following the pick-up, nanobeams are cantilevered on the edge of an oxidised Si chip and a miniscule amount of very low viscosity, UV-curable epoxy is applied to the end of the nanobeam on the substrate. Once the epoxy is cured, nanobeam is securely attached to the substrate. After curing, the sample is dipped into BOE solution one more time to get rid of any surface oxide layer that might have formed on the nanobeam and any contaminants that might prevent Pd coverage. The Pd layer is evaporated straight from below to cover the tip of the nanobeam as shown in Fig. S2a.

When preparing nanobeams on PDMS substrate, just like the cantilevered beams we first loosen the crystals from the growth surface with BOE. Then, an oxidised Si chip spin-coated with PDMS is cured and placed onto the chip with loosened nanobeams. Once removed, some $VO_2$ nanobeams stick to PDMS surface. Before evaporating the Pd layer, samples are diffed in BOE for a couple of seconds for the reasons mentioned earlier.

## 3. Further Evidence for Anisotropic Diffusion

In Figure 2c of the main text, a common observation is shown; while warming hydrogen doped nanobeams, we see distinct regions with different levels of darkening. Fig. S3a provides another evidence for our explanation given in the main text. Fig. S3a shows two optical microscope images of the same nanoplate, the image on the left shows the top surface and the image on the right shows the bottom side of the nanobeam, taken at the room temperature. The Pd layer is evaporated from above and the nanoplate is exposed to hydrogen at 120 °C for 15 minutes. As indicated by small arrows on the images, clearly there is a difference in the levels of darkness at the ends of the nanobeam before and after flipping the crystal upside down.

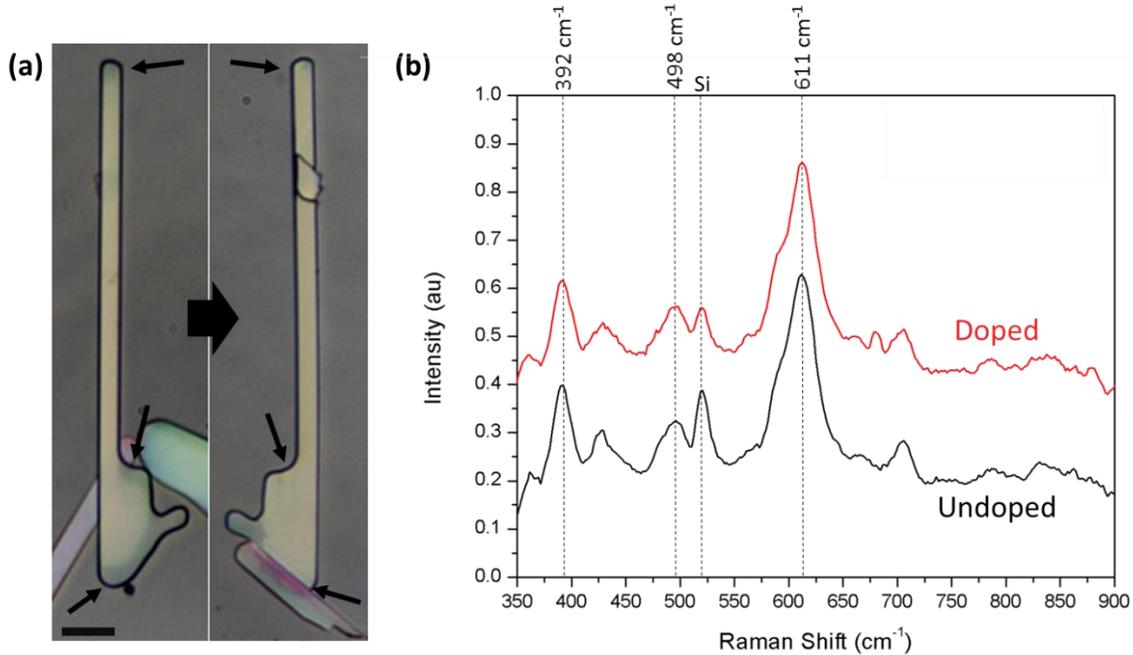

**Figure S3**. **(a)** Images of the top (left image) and the bottom faces of a hydrogenated nanobeam. The Pd layer is evaporated to top face of the crystal directly from above. As indicated by arrows it is evident from the color that the region close to the top face got doped much more than the face close to the bottom face. Scale bar is 5 μm. **(b)** Room temperature Raman spectra of a $VO_2$ crystal before and after mild doping at 100 °C for 30 mins. Excitation laser is placed very close to the metallic region. There is no observable change in the Raman signal.

## 4. Determining the Diffusion Length, $x_m$

To measure the diffusion coefficient and the activation energy along the rutile c-axis we used nanobeams that exhibit behaviour consistent with uniform hydrogen doping at the ends at moderate exposure levels. Length measurements are taken from digitally captured microscope images using MB-ruler software. Length calibration of the images are performed with an AFM calibration grating. Our observations reveal that as the sample temperature gets closer to the intrinsic critical temperature, the boundary between insulating and metallic phases gets sharper as shown in Fig. S4. In order to track the interface consistently across the pictures, we fixed the exposure level and the white balance of the camera as well as the illumination and the diaphragm of the microscope. We picked a color at the interface and looked for the same RGB value in other images with an image processing software and measured its distance from the tip of the nanobeam. This measurement is taken at 60 °C since the boundary is sharp enough to give a precise location of the interface and gives a good measure of how far hydrogen is diffused. Our measured value for the activated diffusion coefficient gives the correct diffusion lengths for the cantilevered nanobeams as well.

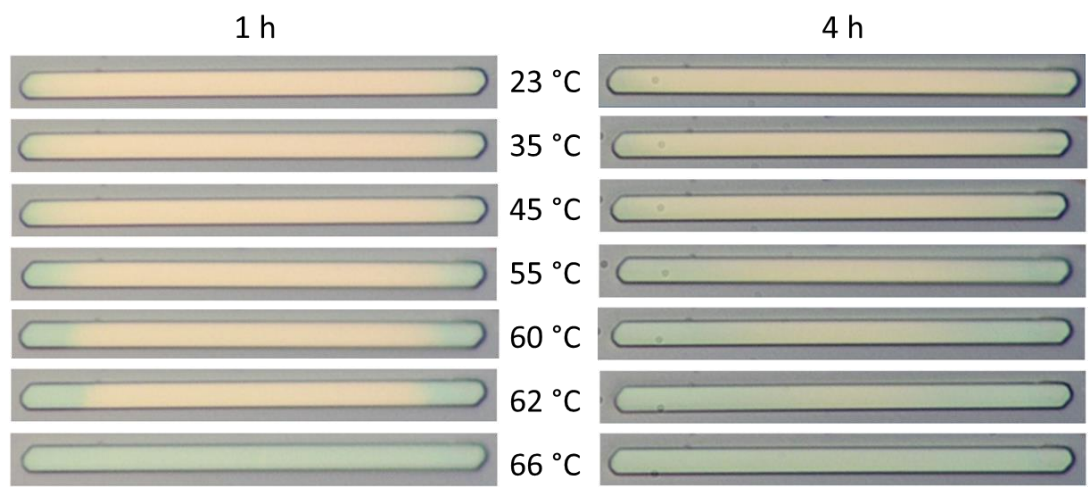

**Figure S4.** Series of pictures of a nanobeam at different stage temperatures, hydrogenated at 100 °C under $H_2$:Ar (1:10) gas mixture flow for 1 hour and 4 hours. Notice how the metallic region becomes sharper as the temperature increase. Just around 65 °C rest of the nanobeam turns metallic. Scale bar is 10 µm.

## 5. Dehydrogenation and Further Evidence for spontaneous segregation

Fig. S5 shows two more pieces of evidence for spontaneous segregation, ie, behavior which is the opposite of the smoothing of the hydrogen profile with time.

Fig. S5a shows a nanobeam hydrogenated for 4 hours at 100 °C under $H_2$:Ar (1:10) gas mixture and Figure S5b shows a series of images of the same nanobeam annealed under pure Ar gas flow at 100 °C, taken after different durations. Since the diffusion is so anisotropic, during the process we expect hydrogen to exit primarily from the ends. The images are taken at 60 °C, at which the contrast between metallic and insulating parts is larger than at room temperature. If the diffusion were linear, the profile would become more and more uniform with time, and the concentration would be lower at the ends as hydrogen exits there. In contrast, after 2-4 hours, we see that the middle becomes paler than the ends, fairly distinct boundaries appear (indicated by arrows), and the ends actually get slightly darker than they were before. After 12 hours there are no signs of hydrogen left in the entire nanobeam, which turns metallic uniformly upon reaching the intrinsic critical temperature of 65 °C. It may be that the segregation actually helps to remove the hydrogen here by pushing it towards the ends.

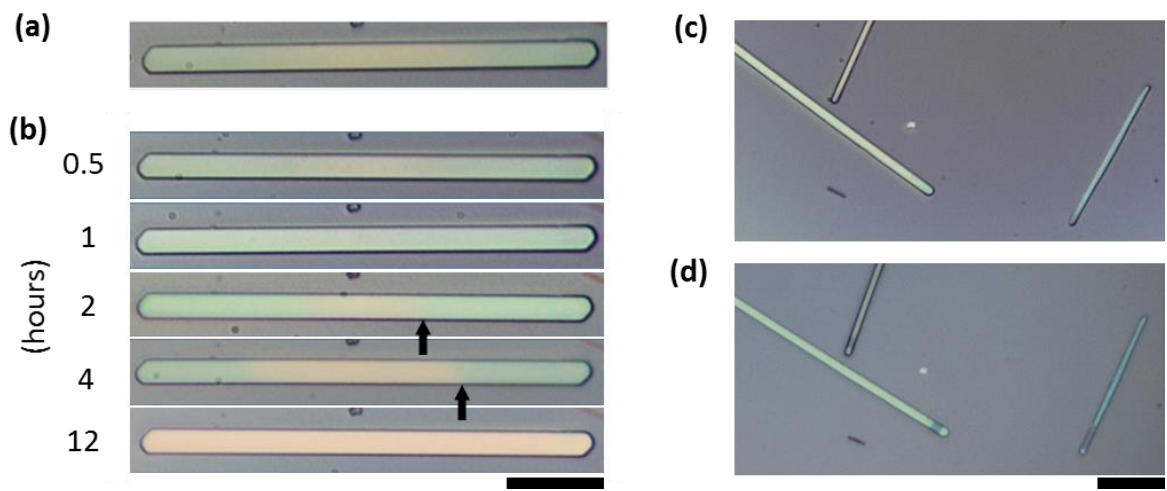

**Figure S5. (a)** VO$_2$ nanobeam after 4 hours of hydrogenation at 100 °C under H$_2$:Ar (1:10) gas mixture flow. **(b)** A series of images of the same nanobeam, annealed under Ar flow (500 mL/min) at 100 °C, taken after different annealing durations. **(c)** VO$_2$ nanobeams hydrogenated at 100 °C for 15 minutes under H$_2$:Ar (1:10) gas mixture flow. **(d)** Same nanobeams after 5 months kept under ambient conditions clearly show sharp metallic stripes several microns away from the tips. All images are taken at 60 °C and the scale bars are 10 μm.

Fig. S5c shows nanobeams with darkened ends after hydrogen exposure at 100 °C, and Fig. S5d shows the same nanobeams after storing at room temperature in air for five months. The hydrogen appears to have accumulated in patches near the ends. This can be interpreted as the self trapping of domains of enhanced H doping: H near the end diffuses away from the end as a result of the segregation tendency, and possibly at room temperature the rate of leaving the ends is much lower than at 100 °C.

We observe formation of metallic stripes in different nanobeams similar to that is shown in Fig. 4d and 4e, in the main text. Fig. S6 shows segregated metallic regions in different nanobeams hydrogenated at different temperatures for varying durations.

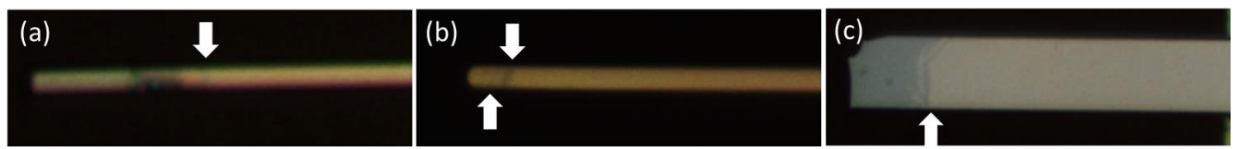

**Figure S6.** Formation of metallic stripes in several different nanobeams are shown. Nanobeams shown in **(a)** and **(b)** are hydrogenated for 3 hours at 90 °C and 70 °C respectively and the nanobeam shown in **(c)** is hydrogenated for 11 minutes at 120 °C. Metallic stripe formation is evident in all three nanobeams as indicated by arrows.

# References

[1] Park, J. H.; Coy, J. M.; Kasirga, T. S.; Huang, C.; Fei, Z.; Hunter, S.; Cobden, D. H. Measurement of a sold-state triple point at the metal-insulator transition in VO$_2$. *Nature* **2013**, *500*, 431-434.